\documentclass[aps,10pt,pra,superscriptaddress,showpacs,floatfix,twocolumn,longbibliography]{revtex4-1}

\usepackage{graphicx}
\usepackage{epstopdf}
\usepackage{amsmath}
\usepackage{comment}
\usepackage{amssymb}
\usepackage{hyperref}
\usepackage{bm}
\usepackage{subfigure}

\hypersetup{
    colorlinks=true,       
    linkcolor=cyan,          
    citecolor=magenta,        
    filecolor=magenta,      
    urlcolor=cyan,           
    runcolor=cyan
}
\usepackage[pdftex]{color}

\newcommand{\beq}{\begin{equation}}
\newcommand{\eeq}{\end{equation}}
\newcommand{\beqnn}{\begin{equation*}}
\newcommand{\eeqnn}{\end{equation*}}
\newcommand{\bea}{\begin{eqnarray}}
\newcommand{\eea}{\end{eqnarray}}
\newcommand{\beann}{\begin{eqnarray*}}
\newcommand{\eeann}{\end{eqnarray*}}
\newcommand{\bes} {\begin{subequations}}
\newcommand{\ees} {\end{subequations}}

\newcommand{\ket}[1]{ | #1\rangle}

\newcommand{\ignore}[1]{}

\newcommand{\rme}{\text e}
\newcommand{\rmd}{{\text d}}

\begin{document}

\title{How Quantum is the Speedup in Adiabatic Unstructured Search?}

\author{Itay Hen}
\affiliation{Information Sciences Institute, University of Southern California, Marina del Rey, California 90292, USA}
\affiliation{Department of Physics and Astronomy and Center for Quantum Information Science \& Technology, University of Southern California, Los Angeles, California 90089, USA}
\email{itayhen@isi.edu}
\begin{abstract}
In classical computing, analog approaches have sometimes appeared to be more powerful than they really are. This occurs when resources, particularly precision, are not appropriately taken into account. While the same should also hold for analog quantum computing, precision issues are often neglected from the analysis. In this work we present a classical analog algorithm for unstructured search that can be viewed as analogous to the quantum adiabatic unstructured search algorithm devised by Roland and Cerf [Phys. Rev. A 65, 042308 (2002)]. We show that similarly to its quantum counterpart, the classical construction may also provide a quadratic speedup over standard digital unstructured search. We discuss the meaning and the possible implications of this result in the context of adiabatic quantum computing.
\end{abstract}

\maketitle
%


\section{Introduction}

Along with Shor's polynomial-time algorithm for integer factorization~\cite{shor:94}, 
Grover's unstructured search algorithm~\cite{grover:97} is considered a tour de force of quantum computing,  exhibiting one of only a few examples to date of the superiority of quantum computers over classical.
Classically, the number of queries required for finding a marked item in an unsorted database scales linearly with the number of elements $N$, while  Grover's quantum circuit requires only $O(\sqrt{N})$ calls. 

A different quantum algorithm for unstructured search yielding the same quadratic speedup has been proposed in the framework of adiabatic quantum computing \cite{Finnila1994343,Brooke30041999,kadowaki:98,farhi:01,santoro:02}---a paradigm of computation that is viewed by many as a simpler way of carrying out quantum-assisted calculations and is easier and perhaps more natural to implement experimentally~\cite{vandersypen:01,gaitan:12,speedup,DWave,Hen:2015rt,q-sig,q-sig2,DWave,QEO,LL1,LL2,LL3,DW2000}. Adiabatic quantum computing is a gate-free method in which a time-evolving Hamiltonian that uses continuously decreasing quantum fluctuations is employed 
to find the global optima of discrete optimization problems in an analog, rather than digital, manner~\cite{young:08,young:10,hen:11,hen:12,farhi:12}.

The quantum adiabatic search algorithm, originally devised by Roland and Cerf~\cite{roland:02} (but see also Refs.~\cite{analogAnalogue,howPowerful} for earlier variants), consists of encoding the search space in a `problem Hamiltonian,' $H_p$, that is constant across the entire search space except for one `marked' configuration $\ket{m}=\ket{m_1 m_2 \ldots m_n}$ whose cost is lower than the rest. Here, $m_i \in \{0,1\}$ are the bits of the $n$-bit solution $\ket{m}$ (the number of elements in the search space is thus $N=2^n$). 

In terms of distinct $k$-body interactions, the unstructured search problem Hamiltonian, which is a one-dimensional projection onto the marked state, decomposes to a sum of $2^n$ terms, namely,
\bea\label{eq:Hf}
H_p &=& - |m\rangle\langle m| \\
&=&-\frac1{2^n} \left( 1+\sum_i (-1)^{m_i} \sigma^z_i +\sum_{i < j} (-1)^{m_i+m_j} \sigma^z_i \sigma^z_j \right. \nonumber\\
&+& \left. \sum_{i < j <k} (-1)^{m_i+m_j+m_k} \sigma^z_i \sigma^z_j \sigma^z_k + \ldots + \prod_i (-1)^{m_i} \sigma^z_i\right)\nonumber \,.
\eea
To achieve the quadratic speedup, a carefully tailored variable-rate annealing schedule is selected which evolves the Hamiltonian in time between a `beginning' Hamiltonian $H_b$ whose purpose is to provide the quantum fluctuations, and $H_p$, such that the state of the system remains close to the instantaneous ground state throughout the evolution,  varying slowly in the vicinity of the minimum gap and allowed to evolve more rapidly in places where the gap is large~\cite{jansen:07,lidarGap,kato:51,roland:02}. Here, the beginning  Hamiltonian $H_b$ is a one-dimensional projection onto the equal superposition of all computational basis states, i.e., $H_b=-|+\rangle \langle+|$ and the total Hamiltonian is given by
\beq\label{eq:hs}
{H}(s)=(1-s) H_b + s H_p \,,
\eeq
 where $s(t)$ is the annealing schedule which varies smoothly with time $t$
from $s(0)=0$ initially to $s(\mathcal{T})=1$ at the end of the evolution. 

The quantum adiabatic search algorithm is nonstandard in several ways. First, the one-dimensional projection Hamiltonian that it uses for an oracle may be viewed as physically unrealizable.  Unlike its circuit-based counterpart~\cite{grover:97}, any implementation of the oracle must contain highly non-local interactions, requiring up to $n$-body terms, and exponentially, many more~\cite{hen:14,hen:14b,realizableAQCsearch}. The same holds for the beginning Hamiltonian. Moreover, the quantum adiabatic algorithm is purely analog in nature, requiring continuously varying coupling strengths throughout the evolution~\cite{roland:02,roland:03,hen:14,hen:14b}. Explicitly, the quantum adiabatic gap can be computed to be $g(s)=\sqrt{1-4 s (1-s)(1-1/N)}$ with a minimum gap on the order of $1/\sqrt{N}$ that is centered around $s=1/2$ to within a region of width $1/\sqrt{N}$~\cite{childsGoldstone,realizableAQCsearch}---implying that in order to maintain the quadratic speedup as the problem scales up, an exponentially precise annealing schedule $s(t)$ is required. To wit, the `digitization' of the algorithm, as prescribed by the polynomial equivalence between adiabatic quantum computing and the quantum circuit model~\cite{howPowerful,RevModPhys.90.015002,QAOA}, does not in fact preserve the quadratic speedup, but instead yields a classical $O(N)$ scaling.

Given the above arguments, it is important to ask whether the quadratic speedup produced by the quantum adiabatic Grover algorithm originates from its `quantumness' or is simply a consequence of its analog nature. It is useful to recall that analog computation (be it classical or quantum), in which finite precision is not properly taken into account, may be misleadingly construed as more powerful than it actually is~\cite{Aaronson,VERGIS198691,analogComputing,analogErrors}.

Here, we do not answer this question directly. However we address it by considering the possibility that there exists a classical variant of the quantum adiabatic search algorithm that provides a similar quadratic speedup without utilizing quantum fluctuations. The classical model we propose may in fact be viewed as a direct analog of Roland and Cerf's algorithm, wherein qubits have been replaced with rotors and the Pauli operators in the Hamiltonians are  replaced with expectation values. 

\section{Classical analog unstructured search}

\begin{figure}[ht] 
\includegraphics[width=0.8\columnwidth]{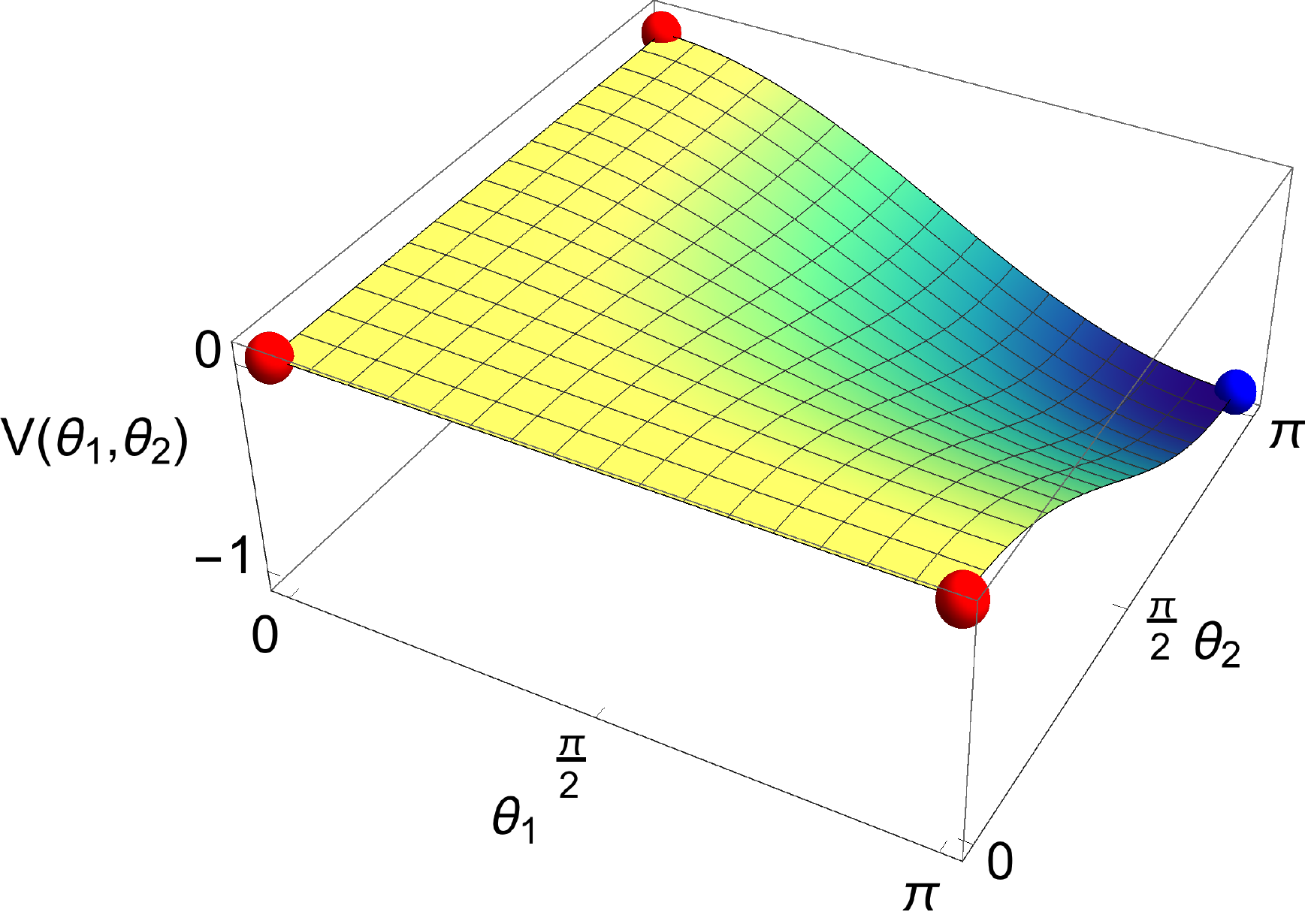}
   \caption{The classical analog potential $V(\theta_1,\theta_2)$ in the two-rotor case. The potential is zero at $(0,0), (0,\pi)$ and $(\pi,0)$, denoted by red spheres, and attains its minimum of $-1$ at $(\pi,\pi)$, denoted by a blue sphere.}
   \label{fig:pot3d}
\end{figure}

Let us consider a system of $n$ two-dimensional rotors, with angular degrees of freedom denoted by $\theta_i$ with \hbox{$i=1\ldots n$}. 
Taking our cue from the quantum adiabatic search algorithm~\cite{roland:02}, we allow the rotors to interact via a similar black-box potential. Starting with the potential given in Eq.~(\ref{eq:Hf}) we demote the Pauli operators $\sigma^z_i$ to their single-qubit expectation values $\cos \theta_i$, ending up with
\bea
V(\theta_i)&=&-\frac{1}{2^n} \left( 1+\sum_i (-1)^{m _i}\cos \theta_i \right. \\\nonumber
&+& \sum_{i<j} (-1)^{m _i+m_j}\cos \theta_i \cos \theta_j \\\nonumber
&+& \left.\sum_{i<j<k} (-1)^{m _i+m_j+m_k}\cos \theta_i \cos \theta_j\cos \theta_k + \ldots \right)
\eea 
The above potential attains its minimum value of $-1$ at \hbox{$\theta^{(\min)}_i=m_i \pi $} (that is, the angle is zero if $m_i=0$, and $\pi$ if $m_i=1$) and vanishes for any other angle combination in $\{0, \pi\}^n$.  
The potential $V$ is plotted in Fig.~\ref{fig:pot3d} for the simple case of $n=2$ with a solution at $(\pi,\pi)$.

Next, the kinetic energy of the rotors is given by $T=\frac1{2} I \sum_i \dot{\theta}^2$ where $I$ is a rotor's moment of inertia (we will choose $I=1$ in the appropriate units). 
Treating $T$ as a source of fluctuations, we consider the following Lagrangian, which interpolates between the kinetic and potential terms: 
\beq
L=[1-s(t)] T- s(t) V
\eeq
 where $s(t)$ is the classical annealing parameter obeying $s(0)=0$ and $s(\mathcal{T})=1$~\footnote{The Lagrangian can be interpreted as a system of rotors with time-varying potential and moments of inertia.}.

We now ask: how long would it take for the rotors to align into a configuration that minimizes the potential? To answer the question, we set up the initial conditions for the angles and their derivatives such that at $t=0$, the state of the system minimizes the kinetic energy $T$, namely, $\dot{\theta}_i=0$ and $\theta_i=\pi/2$ for all $i$ (while the latter condition is not strictly necessary for the minimization of the energy, it is chosen so that the angles favor neither zero nor $\pi$ in the beginning~\footnote{One could consider a small addition to the kinetic term that would remove the degeneracy of its minima and enforce the $\theta_i=\pi/2$ condition.}).

\begin{figure}[ht] 
\includegraphics[width=0.8\columnwidth]{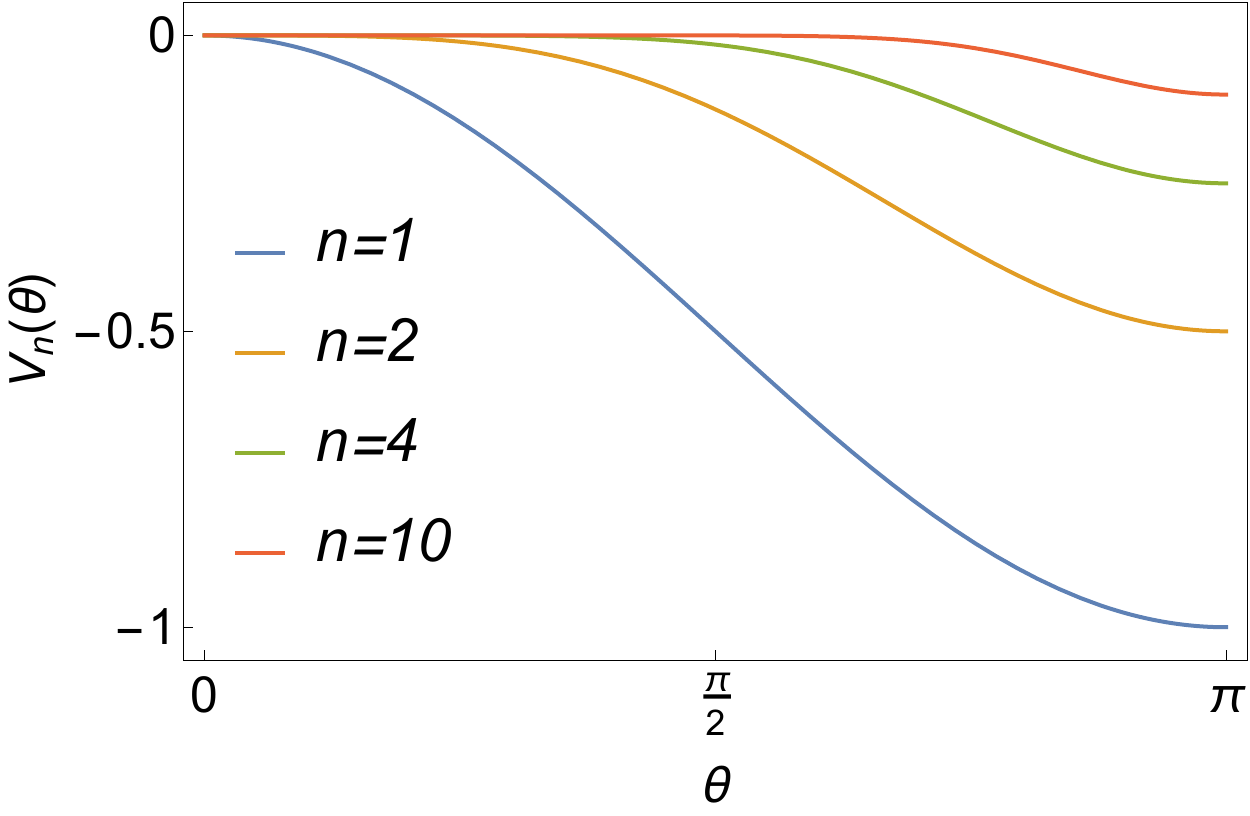}
   \caption{Effective potential $V_n(\theta)$ for the classical analog unstructured search algorithm as a function of rotor angle 
$\theta$ for several system sizes $n$. }
   \label{fig:pot}
\end{figure}

The Euler-Lagrange equations of motion for the system are
\beq
\frac{\rmd}{\rmd t} \frac{\partial L}{\partial \dot{\theta}_i}- \frac{\partial L}{\partial \theta_i}=0 \,.
\eeq
The transformation $\theta_i \to \pi -\theta_i$ for all $i$ for which $m_i=0$ simplifies matters, sending the marked state into the all-$\pi$ solution, and along the way transforming all the equations of motion to the set:
\beq
(1-s)\ddot{\theta}_i-\dot{s} \dot{\theta}_i-\frac1{2}s \sin \theta_i \prod_{j \neq i} \sin^2 \frac{\theta_j}{2}=0 \quad \forall i \,.
\eeq
Since the transformed angles $\theta_i$ all evolve in time in an identical manner, the above set of equations may be reduced to a single one by switching to a single variable $\theta_i \to \theta$, yielding:
\beq\label{eq:1d}
(1-s)\ddot{\theta}-\dot{s} \dot{\theta}-s   \sin^{2 n-1} \frac{\theta}{2}\cos \frac{\theta}{2}=0
\eeq
Equation~(\ref{eq:1d}) describes the evolution of a single rotor interpolating between a kinetic term and an effective potential term of the form
\beq 
V_n(\theta)=-\frac1{n} \sin^{2 n} \frac{\theta}{2}\,.
\eeq
This effective potential is plotted in Fig.~\ref{fig:pot} for various values of $n$. An interesting property of the potential is that it contains no barriers, nor does it have plateaus. It points directly at the correct solution $\theta=\pi$ despite becoming more and more shallow with increasing $n$. 
We thus find that the oracular implementation of the potential provides the search space with an important structure. 

The equation of motion, Eq.~(\ref{eq:1d}), unfortunately has no closed-form solution, making it difficult to find the optimal schedule $s(t)$ that would minimize the time to reach $\theta(\mathcal{T})=\pi$ from the initial $\theta(0)=\pi/2$. 
Nonetheless, a hint as to the performance of the analog algorithm may be gained by studying the special case $s(t)=1/2$~\footnote{In this case, the algorithm is reminiscent of the analog quantum search algorithm proposed in Ref.~\cite{analogAnalogue}.}.
In this case, energy is conserved: 
\beq
E=\frac1{2} \dot{\theta}^2  -\frac1{n} \sin^{2 n} \frac{\theta}{2} = -\frac1{n} \sin^{2 n} \frac{\pi}{4} = -\frac{1}{n 2^n}
\eeq
and its equation can be integrated to give:
\bea
\mathcal{T}&=&\int_0^\mathcal{T} \rmd t =\int_{\pi/2}^{\pi} \rmd \theta \sqrt{\frac{n 2^n}{2^{n+1}\sin^{2n} \frac{\theta}{2} -2}}\\\nonumber
&\approx&\int_{\pi/2}^{\pi} \rmd \theta \sqrt{\frac{n 2^n}{2^{n+1}\sin^{2n} \frac{\theta}{2}}}=\sqrt{n}\,{}_2F_1(\frac1{2},\frac{n+1}{2};\frac{3}{2};\frac1{2})\,,
\eea
where ${}_2F_1$ is the ordinary hypergeometric function. In the large $n$ limit, the above expression simplifies to 
\beq
\mathcal{T}  \approx \sqrt{\frac{2^{n+1}}{n}} =\sqrt{\frac{2}{\log_2 N}} \sqrt{N} \,,
\eeq
yielding, asymptotically, a square root scaling with the size of the search space (up to logarithmic corrections). 

\begin{figure}[htp] 
\includegraphics[width=0.85\columnwidth]{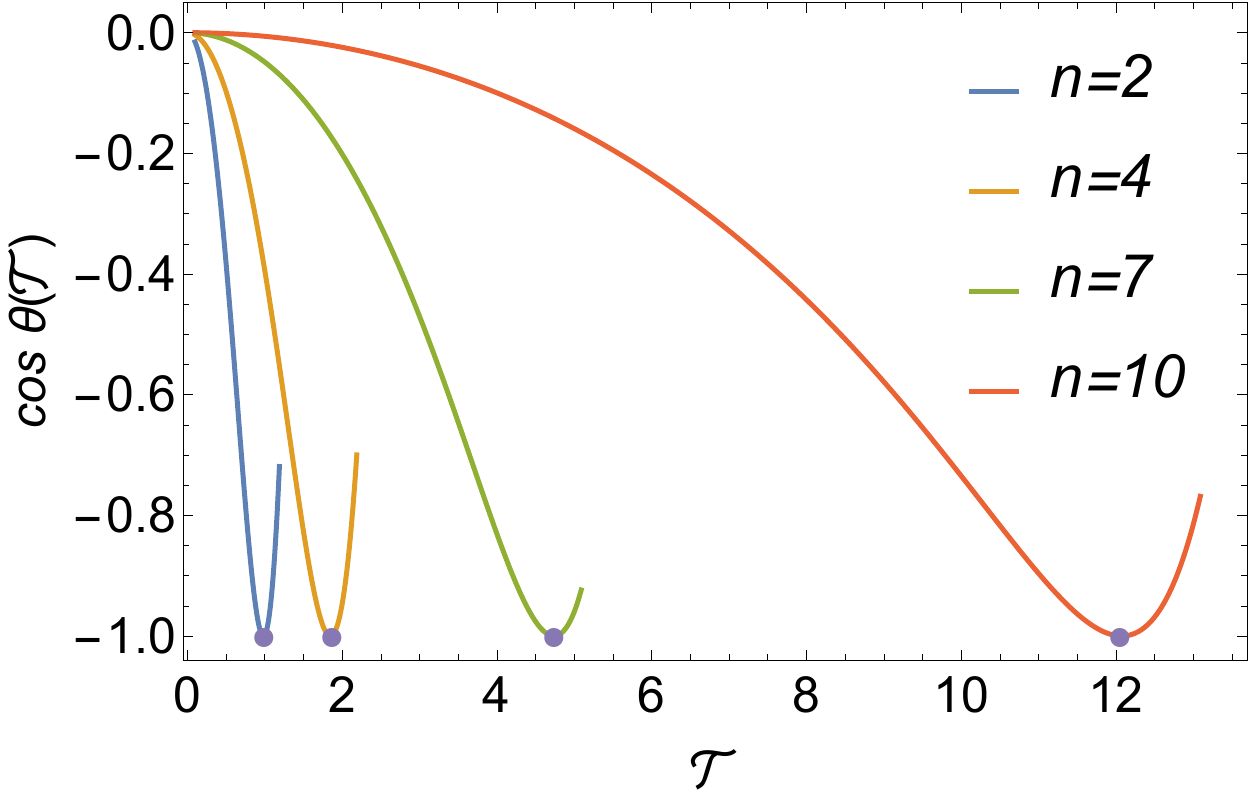}
\includegraphics[width=0.85\columnwidth]{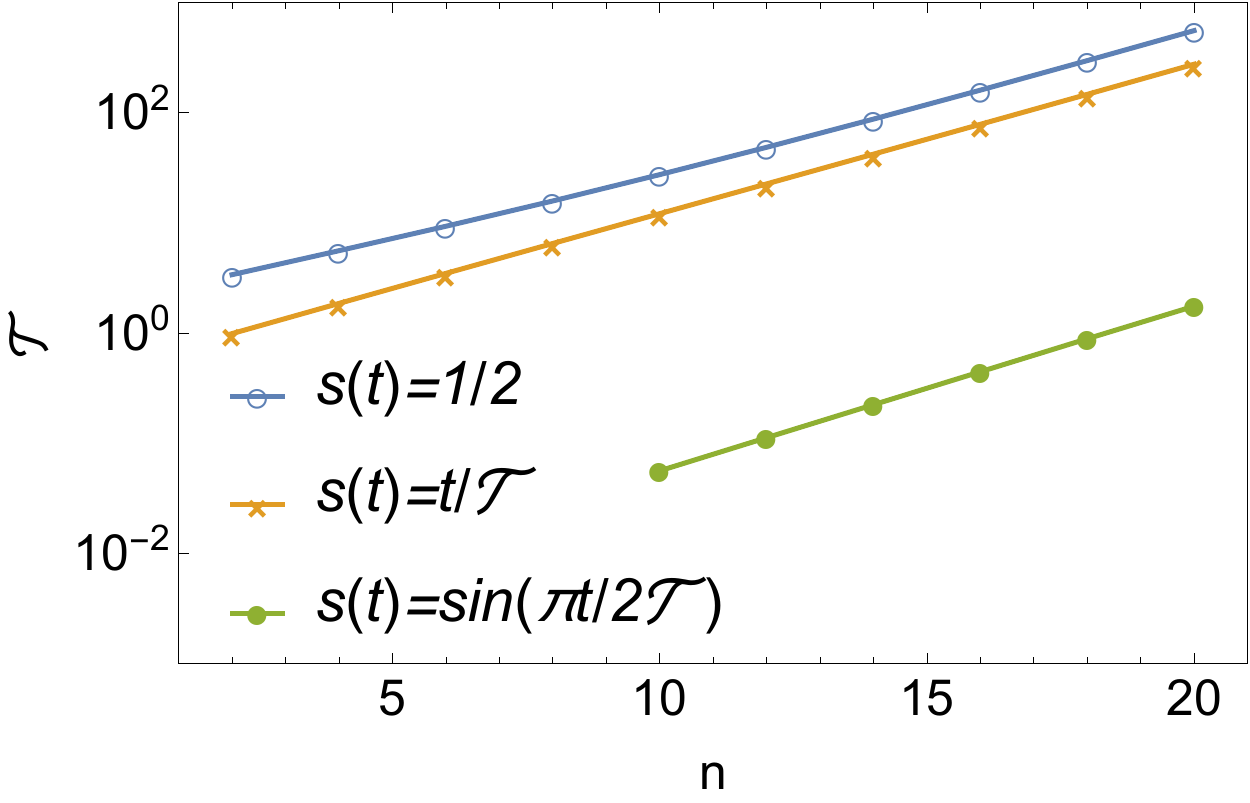}
   \caption{Top: Final rotor angle as a function of runtime $\mathcal{T}$ for different system sizes for a linear schedule $s(t)=t/\mathcal{T}$. The optimal runtime is the first occurrence of $\theta(\mathcal{T})=\pi$, or $\cos \theta(\mathcal{T})=-1$. These occurrences are marked with dots in the figure. Bottom: Optimal runtime $\mathcal{T}$ as a function of problem size $n$ for three different annealing schedules $s(t)$. All schedules show a quadratic speedup, i.e., a scaling of $\sim O(2^{n/2})$.}
   \label{fig:numerics}
\end{figure}

A similar runtime scaling is obtained numerically for two other `standard' schedules, namely, $s(t)=t/\mathcal{T}$ and for $s(t)=\sin \frac{\pi t}{2\mathcal{T}}$. Here, the runtimes are defined as the minimal annealing runtime $\mathcal{T}$ for which $\theta(\mathcal{T})=\pi$, equivalently, $\cos \theta(\mathcal{T})=-1$ [an example is given in Fig.~\ref{fig:numerics} (top) for a linear schedule]. The minimal runtimes for the above schedules, as a function of size, are depicted in Fig.~\ref{fig:numerics}~(bottom), all producing a scaling of $O(2^{n/2})=O(\sqrt{N})$ (with perhaps additional logarithmic corrections).

It would be instructive to note at this point that the potential governing of the evolution of the rotors does not only have structure, but is also devoid of local minima, a property that in turn leads to a seemingly super-classical performance. Explicitly, the torque it exerts on the rotors always points them towards its minimum. The torque on the $r$-th rotor is given by
\bea
\tau_ r&=& - \frac{\partial V}{\partial \theta_r}=-\frac{(-1)^{m _r}\sin \theta_r}{2^n} \left(1+ \sum_{j\neq r} (-1)^{m_j}\cos \theta_j \right. \nonumber\\
&+& \sum_{i \neq r<j\neq r} (-1)^{m _i+m_j}\cos \theta_i \cos \theta_j \\\nonumber
&+& \left.\sum_{i \neq r <j \neq r<k \neq r} (-1)^{m _i+m_j+m_k}\cos \theta_i \cos \theta_j\cos \theta_k + \ldots \right) \,.
\eea
Since the term in parenthesis is always negative [it is in fact the potential term for a system of ($n-1$) rotors] the directionality, or sign, of the torque is directly proportional to $(-1)^{m_r}$, the bit that sets its minimizing angle. Determining the minimizing configuration of the potential thus requires only determining the direction in which each of the rotors rotates. Interestingly, the ability to do so hinges on the sensitivity, or resolving power, of the observing device (be it the eye or some type of device). Since the angular resolution of any such device cannot be assumed to scale with system size, the time required to discern the direction in which the rotors move is linearly proportional to the time it takes the rotors to rotate a full quadrant to hit the marked state. 

\section{Effects of finite precision}

So far, the role that precision requirements play in the performance of the proposed algorithm has not been discussed explicitly. One may ask whether it is indeed the hidden infinite precision that characterizes analog computation that allows for the speedup over digital search.  
To that aim, we next study the effect of having limited precision on performance by examining the probability of success to find the marked state at the end of the run in the presence small perturbations, or errors, in the initial conditions. Specifically, we consider a deviation $\epsilon$ in the angles of the rotors at the beginning of the run, namely $\theta_i(0) \to \pi/2+\epsilon$, where $\epsilon$ is a small constant. For simplicity, we assume the same error $\epsilon$ for all angles and consider the schedule $s(t)=1/2$. We note though that the effects on performance are expected to be similar, if not more pronounced, for time-varying schedules or in the presence of uncorrelated errors. 

\begin{figure}[htp] 
\includegraphics[width=0.85\columnwidth]{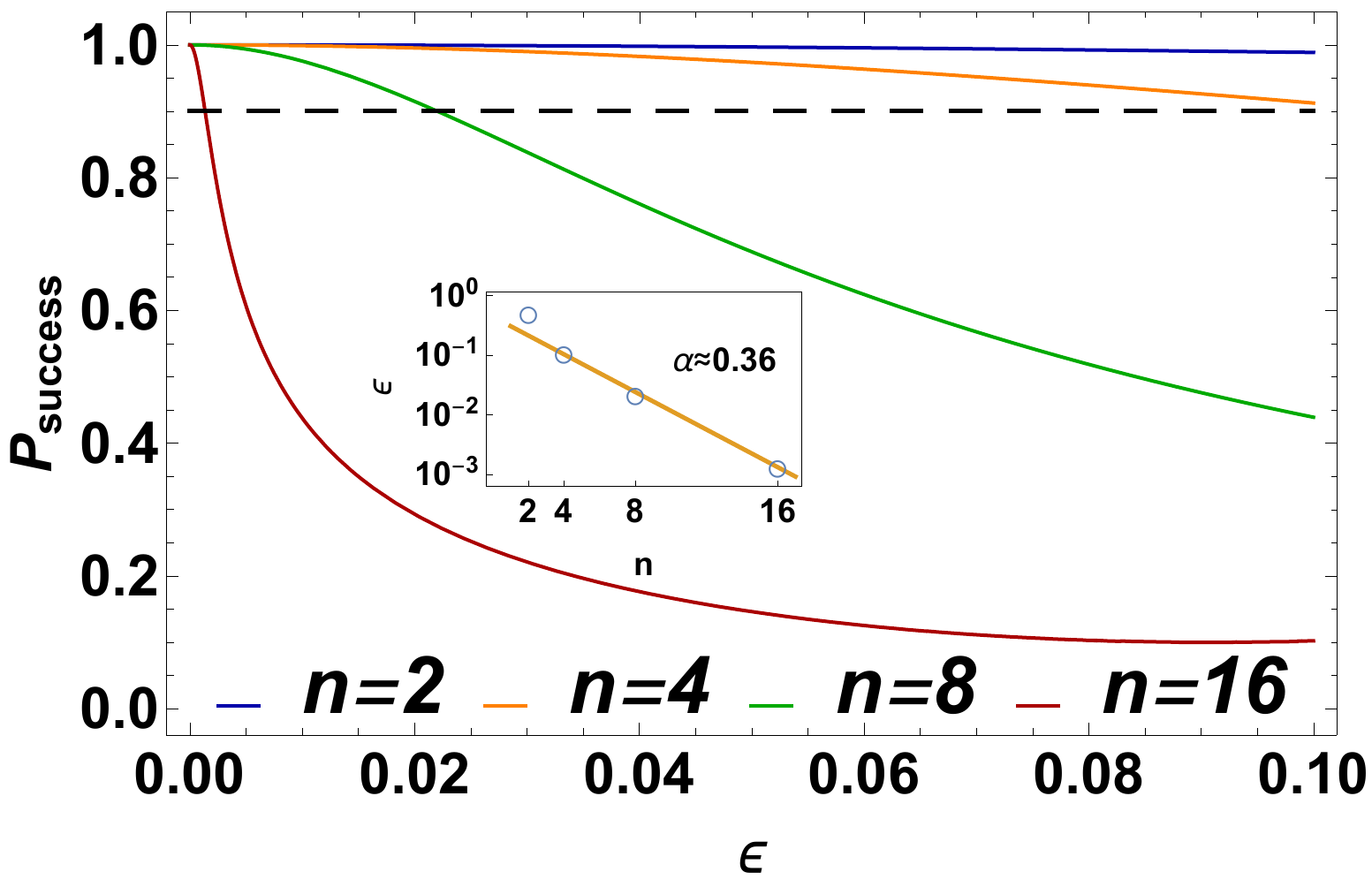}
   \caption{Probability of success $P_{\text{success}}$ as a function of misspecification in the initial angle $\epsilon$ for different system sizes $n$. As $n$ grows, the performance of the algorithm becomes more and more sensitive to perturbations in the initial conditions. Inset: The value of the error $\epsilon$ at which  $P_{\text{success}}=0.9$ (dashed line in main panel) as a function of $n$ (log-linear scale). A fit to $\rme^{-\alpha n}$ is also shown.}
   \label{fig:sensitivity}
\end{figure}

The results are summarized in Fig.~\ref{fig:sensitivity} which depicts the probability of success of the algorithm, $P_{\text{success}}$, as a function of the initial angle misspecification $\epsilon$ for different system sizes. As is evident from the figure, the performance of the algorithm is more adversely affected by the perturbation $\epsilon$ as system size grows. We can quantify this effect by recording the value of $\epsilon$ for which the probability of success drops by some fixed amount, e.g., $P_{\text{success}}=0.9$ as a function of system size. This is shown in the inset of Fig.~\ref{fig:sensitivity}. We find that the allowed error decays exponentially with system size, requiring increasing precision if the quadratic speedup over digital search is to be maintained. 

\section{Conclusions and discussion}

In this study we provided an example of a classical analog construction, inspired by the quantum adiabatic unstructured search algorithm devised by Roland and Cerf~\cite{roland:02}, for searching an unstructured database. We showed that the required runtime for finding the marked item scales as the square root of the number of elements in the search space, similar to the analogous quantum construction. Since the classical algorithm presented here does not possess any uniquely quantum properties such as entanglement or massive parallelism, the possibility arises that the speedup of the quantum adiabatic search algorithm is a result of it being analog rather than quantum.  

We also demonstrated that the classical algorithm is indeed sensitive to small perturbations in the setup, specifically, in initial conditions, as analog algorithms typically are~\cite{Aaronson,VERGIS198691,analogComputing,analogErrors}. We found that the algorithm's precision requirements increase with problem size if its quadratic speedup is to be maintained.  
In this context, it is also useful to note again that obtaining the quantum speedup for the Roland and Cerf algorithm requires an `exponentially precise' annealing schedule, as the minimum gap is exponentially localized~\cite{childsGoldstone,realizableAQCsearch,Jonckheere2013}, and that further digitization of the algorithm into a circuit by Trotterization~\cite{howPowerful,RevModPhys.90.015002,QAOA} does not preserve the quadratic quantum speedup~\cite{analogQAUS}. Nonetheless, it should be noted that newer algorithms for simulating Hamiltonian evolutions using quantum circuits offer more efficient digitization techniques~\cite{Berry,Somma1,Somma2,Somma3}. These, however, primarily apply to sparse Hamiltonians. It would be interesting to see whether and how much of the quantum advantage of the Roland and Cerf algorithm is sustained if one employs these advanced techniques.

The classical analog construction proposed in this work shares several other key similarities with its quantum adiabatic counterpart. First, in the present algorithm one takes advantage of the symmetries of the problem in order to show that the evolution takes place in a sub-manifold of an exponentially reduced dimensionality, while in the quantum case, one could write down an effective two-by-two Hamiltonian to describe the evolution of the spins in the system. In the classical case the evolution of the two-dimensional rotors is described using an equivalent second-order differential equation. Second, the oracular potential in the present algorithm interpolates between the originally discretized search space states, accepting as input superpositions of digital queries. This property in turn provides the potential for an all-important structure. As we have demonstrated, the added structure to an otherwise unstructured search space is eventually translated to a runtime speedup over the digital unstructured case. One may therefore wonder whether the oracles in both the adiabatic quantum and the analog classical cases are somehow more powerful than intended---especially given their physical infeasibility which may be playing an unintentionally important role.

It should be explicitly noted perhaps that the classical construction proposed here is not meant to be experimentally realizable but was conceived to highlight the potential pitfalls of analog --- in this case quantum --- algorithms. In light of the results presented here, it would be of interest to therefore rigorously determine whether the quantum adiabatic quadratic speedup for unstructured search is not indeed a consequence of the infinite precision possessed by ideal analog computation (or a too powerful oracle). Of course, similar arguments may also be posed for other quantum analog algorithms. We leave this for future work~\cite{analogQAUS}. 

Finally, it is also worth noting that the issues raised in this study, namely, the need for infinite precision and the infeasibility of the oracle, do not arise in the context of Grover's original gate-based algorithm as the latter is not only digital but the unitary operations required to `mark' the sought state can be efficiently carried out~\cite{nielsen:00}. 

\begin{acknowledgements}
We thank Tameem Albash, Elizabeth Crosson, Daniel Lidar and Eleanor Rieffel for insightful discussions.
The research is based upon work (partially) supported by the Office of
the Director of National Intelligence (ODNI), Intelligence Advanced
Research Projects Activity (IARPA), via the U.S. Army Research Office
contract W911NF-17-C-0050. The views and conclusions contained herein are
those of the authors and should not be interpreted as necessarily
representing the official policies or endorsements, either expressed or
implied, of the ODNI, IARPA, or the U.S. Government. The U.S. Government
is authorized to reproduce and distribute reprints for Governmental
purposes notwithstanding any copyright annotation thereon.
\end{acknowledgements}
\bibliography{refs}

\begin{thebibliography}{51}%
\makeatletter
\providecommand \@ifxundefined [1]{%
 \@ifx{#1\undefined}
}%
\providecommand \@ifnum [1]{%
 \ifnum #1\expandafter \@firstoftwo
 \else \expandafter \@secondoftwo
 \fi
}%
\providecommand \@ifx [1]{%
 \ifx #1\expandafter \@firstoftwo
 \else \expandafter \@secondoftwo
 \fi
}%
\providecommand \natexlab [1]{#1}%
\providecommand \enquote  [1]{``#1''}%
\providecommand \bibnamefont  [1]{#1}%
\providecommand \bibfnamefont [1]{#1}%
\providecommand \citenamefont [1]{#1}%
\providecommand \href@noop [0]{\@secondoftwo}%
\providecommand \href [0]{\begingroup \@sanitize@url \@href}%
\providecommand \@href[1]{\@@startlink{#1}\@@href}%
\providecommand \@@href[1]{\endgroup#1\@@endlink}%
\providecommand \@sanitize@url [0]{\catcode `\\12\catcode `\$12\catcode
  `\&12\catcode `\#12\catcode `\^12\catcode `\_12\catcode `\%12\relax}%
\providecommand \@@startlink[1]{}%
\providecommand \@@endlink[0]{}%
\providecommand \url  [0]{\begingroup\@sanitize@url \@url }%
\providecommand \@url [1]{\endgroup\@href {#1}{\urlprefix }}%
\providecommand \urlprefix  [0]{URL }%
\providecommand \Eprint [0]{\href }%
\providecommand \doibase [0]{http://dx.doi.org/}%
\providecommand \selectlanguage [0]{\@gobble}%
\providecommand \bibinfo  [0]{\@secondoftwo}%
\providecommand \bibfield  [0]{\@secondoftwo}%
\providecommand \translation [1]{[#1]}%
\providecommand \BibitemOpen [0]{}%
\providecommand \bibitemStop [0]{}%
\providecommand \bibitemNoStop [0]{.\EOS\space}%
\providecommand \EOS [0]{\spacefactor3000\relax}%
\providecommand \BibitemShut  [1]{\csname bibitem#1\endcsname}%
\let\auto@bib@innerbib\@empty
\bibitem [{\citenamefont {Shor}(1994)}]{shor:94}%
  \BibitemOpen
  \bibfield  {author} {\bibinfo {author} {\bibfnamefont {P.~W.}\ \bibnamefont
  {Shor}},\ }\bibfield  {title} {\enquote {\bibinfo {title} {Algorithms for
  quantum computing: discrete logarithms and factoring},}\ }in\ \href@noop {}
  {\emph {\bibinfo {booktitle} {Proc. 35th Symp.~on Foundations of Computer
  Science}}},\ \bibinfo {editor} {edited by\ \bibinfo {editor} {\bibfnamefont
  {S.}~\bibnamefont {Goldwasser}}}\ (\bibinfo {year} {1994})\ p.\ \bibinfo
  {pages} {124}\BibitemShut {NoStop}%
\bibitem [{\citenamefont {Grover}(1997)}]{grover:97}%
  \BibitemOpen
  \bibfield  {author} {\bibinfo {author} {\bibfnamefont {L.~K.}\ \bibnamefont
  {Grover}},\ }\bibfield  {title} {\enquote {\bibinfo {title} {Quantum
  mechanics helps in searching for a needle in a haystack},}\ }\href@noop {}
  {\bibfield  {journal} {\bibinfo  {journal} {Phys. Rev. Lett.}\ }\textbf
  {\bibinfo {volume} {79}},\ \bibinfo {pages} {325} (\bibinfo {year}
  {1997})}\BibitemShut {NoStop}%
\bibitem [{\citenamefont {Finnila}\ \emph {et~al.}(1994)\citenamefont
  {Finnila}, \citenamefont {Gomez}, \citenamefont {Sebenik}, \citenamefont
  {Stenson},\ and\ \citenamefont {Doll}}]{Finnila1994343}%
  \BibitemOpen
  \bibfield  {author} {\bibinfo {author} {\bibfnamefont {A.B.}\ \bibnamefont
  {Finnila}}, \bibinfo {author} {\bibfnamefont {M.A.}\ \bibnamefont {Gomez}},
  \bibinfo {author} {\bibfnamefont {C.}~\bibnamefont {Sebenik}}, \bibinfo
  {author} {\bibfnamefont {C.}~\bibnamefont {Stenson}}, \ and\ \bibinfo
  {author} {\bibfnamefont {J.D.}\ \bibnamefont {Doll}},\ }\bibfield  {title}
  {\enquote {\bibinfo {title} {Quantum annealing: A new method for minimizing
  multidimensional functions},}\ }\href {\doibase
  http://dx.doi.org/10.1016/0009-2614(94)00117-0} {\bibfield  {journal}
  {\bibinfo  {journal} {Chemical Physics Letters}\ }\textbf {\bibinfo {volume}
  {219}},\ \bibinfo {pages} {343 -- 348} (\bibinfo {year} {1994})}\BibitemShut
  {NoStop}%
\bibitem [{\citenamefont {Brooke}\ \emph {et~al.}(1999)\citenamefont {Brooke},
  \citenamefont {Bitko}, \citenamefont {F.}, \citenamefont {Rosenbaum},\ and\
  \citenamefont {Aeppli}}]{Brooke30041999}%
  \BibitemOpen
  \bibfield  {author} {\bibinfo {author} {\bibfnamefont {J.}~\bibnamefont
  {Brooke}}, \bibinfo {author} {\bibfnamefont {D.}~\bibnamefont {Bitko}},
  \bibinfo {author} {\bibfnamefont {T.}~\bibnamefont {F.}}, \bibinfo {author}
  {\bibnamefont {Rosenbaum}}, \ and\ \bibinfo {author} {\bibfnamefont
  {G.}~\bibnamefont {Aeppli}},\ }\bibfield  {title} {\enquote {\bibinfo {title}
  {Quantum annealing of a disordered magnet},}\ }\href {\doibase
  10.1126/science.284.5415.779} {\bibfield  {journal} {\bibinfo  {journal}
  {Science}\ }\textbf {\bibinfo {volume} {284}},\ \bibinfo {pages} {779--781}
  (\bibinfo {year} {1999})}\BibitemShut {NoStop}%
\bibitem [{\citenamefont {Kadowaki}\ and\ \citenamefont
  {Nishimori}(1998)}]{kadowaki:98}%
  \BibitemOpen
  \bibfield  {author} {\bibinfo {author} {\bibfnamefont {T.}~\bibnamefont
  {Kadowaki}}\ and\ \bibinfo {author} {\bibfnamefont {H.}~\bibnamefont
  {Nishimori}},\ }\bibfield  {title} {\enquote {\bibinfo {title} {Quantum
  annealing in the transverse {I}sing model},}\ }\href@noop {} {\bibfield
  {journal} {\bibinfo  {journal} {Phys. Rev. E}\ }\textbf {\bibinfo {volume}
  {58}},\ \bibinfo {pages} {5355} (\bibinfo {year} {1998})}\BibitemShut
  {NoStop}%
\bibitem [{\citenamefont {Farhi}\ \emph {et~al.}(2001)\citenamefont {Farhi},
  \citenamefont {Goldstone}, \citenamefont {Gutmann}, \citenamefont {Lapan},
  \citenamefont {Lundgren},\ and\ \citenamefont {Preda}}]{farhi:01}%
  \BibitemOpen
  \bibfield  {author} {\bibinfo {author} {\bibfnamefont {E.}~\bibnamefont
  {Farhi}}, \bibinfo {author} {\bibfnamefont {J.}~\bibnamefont {Goldstone}},
  \bibinfo {author} {\bibfnamefont {S.}~\bibnamefont {Gutmann}}, \bibinfo
  {author} {\bibfnamefont {J.}~\bibnamefont {Lapan}}, \bibinfo {author}
  {\bibfnamefont {A.}~\bibnamefont {Lundgren}}, \ and\ \bibinfo {author}
  {\bibfnamefont {D.}~\bibnamefont {Preda}},\ }\bibfield  {title} {\enquote
  {\bibinfo {title} {A quantum adiabatic evolution algorithm applied to random
  instances of an {NP}-complete problem},}\ }\href@noop {} {\bibfield
  {journal} {\bibinfo  {journal} {Science}\ }\textbf {\bibinfo {volume}
  {292}},\ \bibinfo {pages} {472} (\bibinfo {year} {2001})}\BibitemShut
  {NoStop}%
\bibitem [{\citenamefont {Santoro}\ \emph {et~al.}(2002)\citenamefont
  {Santoro}, \citenamefont {Marto\v{n}\'ak}, \citenamefont {Tosatti},\ and\
  \citenamefont {Car}}]{santoro:02}%
  \BibitemOpen
  \bibfield  {author} {\bibinfo {author} {\bibfnamefont {G.}~\bibnamefont
  {Santoro}}, \bibinfo {author} {\bibfnamefont {R.}~\bibnamefont
  {Marto\v{n}\'ak}}, \bibinfo {author} {\bibfnamefont {E.}~\bibnamefont
  {Tosatti}}, \ and\ \bibinfo {author} {\bibfnamefont {R.}~\bibnamefont
  {Car}},\ }\bibfield  {title} {\enquote {\bibinfo {title} {Theory of quantum
  annealing of an {I}sing spin glass},}\ }\href@noop {} {\bibfield  {journal}
  {\bibinfo  {journal} {Science}\ }\textbf {\bibinfo {volume} {295}},\ \bibinfo
  {pages} {2427} (\bibinfo {year} {2002})}\BibitemShut {NoStop}%
\bibitem [{\citenamefont {Vandersypen}\ \emph {et~al.}(2001)\citenamefont
  {Vandersypen}, \citenamefont {Steffen}, \citenamefont {Breyta}, \citenamefont
  {Yannoni}, \citenamefont {Sherwood},\ and\ \citenamefont
  {Chuang}}]{vandersypen:01}%
  \BibitemOpen
  \bibfield  {author} {\bibinfo {author} {\bibfnamefont {L.~M.~K.}\
  \bibnamefont {Vandersypen}}, \bibinfo {author} {\bibfnamefont
  {M.}~\bibnamefont {Steffen}}, \bibinfo {author} {\bibfnamefont
  {G.}~\bibnamefont {Breyta}}, \bibinfo {author} {\bibfnamefont {C.~S.}\
  \bibnamefont {Yannoni}}, \bibinfo {author} {\bibfnamefont {M.~H.}\
  \bibnamefont {Sherwood}}, \ and\ \bibinfo {author} {\bibfnamefont {I.~L.}\
  \bibnamefont {Chuang}},\ }\bibfield  {title} {\enquote {\bibinfo {title}
  {Experimental realization of shor’s quantum factoring algorithm using
  nuclear magnetic resonance},}\ }\href@noop {} {\bibfield  {journal} {\bibinfo
   {journal} {Nature}\ }\textbf {\bibinfo {volume} {414}},\ \bibinfo {pages}
  {883–887} (\bibinfo {year} {2001})}\BibitemShut {NoStop}%
\bibitem [{\citenamefont {Bian}\ \emph {et~al.}(2012)\citenamefont {Bian},
  \citenamefont {Chudak}, \citenamefont {Macready}, \citenamefont {Clark},\
  and\ \citenamefont {Gaitan}}]{gaitan:12}%
  \BibitemOpen
  \bibfield  {author} {\bibinfo {author} {\bibfnamefont {Z.}~\bibnamefont
  {Bian}}, \bibinfo {author} {\bibfnamefont {F.}~\bibnamefont {Chudak}},
  \bibinfo {author} {\bibfnamefont {W.~G.}\ \bibnamefont {Macready}}, \bibinfo
  {author} {\bibfnamefont {L.}~\bibnamefont {Clark}}, \ and\ \bibinfo {author}
  {\bibfnamefont {F.}~\bibnamefont {Gaitan}},\ }\bibfield  {title} {\enquote
  {\bibinfo {title} {Experimental determination of ramsey numbers with quantum
  annealing},}\ }\href@noop {} {\  (\bibinfo {year} {2012})},\ \bibinfo {note}
  {(arXiv:1201.1842)}\BibitemShut {NoStop}%
\bibitem [{\citenamefont {R{\o}nnow}\ \emph {et~al.}(2014)\citenamefont
  {R{\o}nnow}, \citenamefont {Wang}, \citenamefont {Job}, \citenamefont
  {Boixo}, \citenamefont {Isakov}, \citenamefont {Wecker}, \citenamefont
  {Martinis}, \citenamefont {Lidar},\ and\ \citenamefont {Troyer}}]{speedup}%
  \BibitemOpen
  \bibfield  {author} {\bibinfo {author} {\bibfnamefont {Troels~F.}\
  \bibnamefont {R{\o}nnow}}, \bibinfo {author} {\bibfnamefont {Zhihui}\
  \bibnamefont {Wang}}, \bibinfo {author} {\bibfnamefont {Joshua}\ \bibnamefont
  {Job}}, \bibinfo {author} {\bibfnamefont {Sergio}\ \bibnamefont {Boixo}},
  \bibinfo {author} {\bibfnamefont {Sergei~V.}\ \bibnamefont {Isakov}},
  \bibinfo {author} {\bibfnamefont {David}\ \bibnamefont {Wecker}}, \bibinfo
  {author} {\bibfnamefont {John~M.}\ \bibnamefont {Martinis}}, \bibinfo
  {author} {\bibfnamefont {Daniel~A.}\ \bibnamefont {Lidar}}, \ and\ \bibinfo
  {author} {\bibfnamefont {Matthias}\ \bibnamefont {Troyer}},\ }\bibfield
  {title} {\enquote {\bibinfo {title} {Defining and detecting quantum
  speedup},}\ }\href {\doibase 10.1126/science.1252319} {\bibfield  {journal}
  {\bibinfo  {journal} {Science}\ }\textbf {\bibinfo {volume} {345}},\ \bibinfo
  {pages} {420--424} (\bibinfo {year} {2014})}\BibitemShut {NoStop}%
\bibitem [{\citenamefont {Johnson}\ \emph {et~al.}(2011)\citenamefont
  {Johnson}, \citenamefont {Amin}, \citenamefont {Gildert}, \citenamefont
  {Lanting}, \citenamefont {Hamze}, \citenamefont {Dickson}, \citenamefont
  {Harris}, \citenamefont {Berkley}, \citenamefont {Johansson}, \citenamefont
  {Bunyk}, \citenamefont {Chapple}, \citenamefont {Enderud}, \citenamefont
  {Hilton}, \citenamefont {Karimi}, \citenamefont {Ladizinsky}, \citenamefont
  {Ladizinsky}, \citenamefont {Oh}, \citenamefont {Perminov}, \citenamefont
  {Rich}, \citenamefont {Thom}, \citenamefont {Tolkacheva}, \citenamefont
  {Truncik}, \citenamefont {Uchaikin}, \citenamefont {Wang}, \citenamefont
  {Wilson},\ and\ \citenamefont {Rose}}]{DWave}%
  \BibitemOpen
  \bibfield  {author} {\bibinfo {author} {\bibfnamefont {M.~W.}\ \bibnamefont
  {Johnson}}, \bibinfo {author} {\bibfnamefont {M.~H.~S.}\ \bibnamefont
  {Amin}}, \bibinfo {author} {\bibfnamefont {S.}~\bibnamefont {Gildert}},
  \bibinfo {author} {\bibfnamefont {T.}~\bibnamefont {Lanting}}, \bibinfo
  {author} {\bibfnamefont {F.}~\bibnamefont {Hamze}}, \bibinfo {author}
  {\bibfnamefont {N.}~\bibnamefont {Dickson}}, \bibinfo {author} {\bibfnamefont
  {R.}~\bibnamefont {Harris}}, \bibinfo {author} {\bibfnamefont {A.~J.}\
  \bibnamefont {Berkley}}, \bibinfo {author} {\bibfnamefont {J.}~\bibnamefont
  {Johansson}}, \bibinfo {author} {\bibfnamefont {P.}~\bibnamefont {Bunyk}},
  \bibinfo {author} {\bibfnamefont {E.~M.}\ \bibnamefont {Chapple}}, \bibinfo
  {author} {\bibfnamefont {C.}~\bibnamefont {Enderud}}, \bibinfo {author}
  {\bibfnamefont {J.~P.}\ \bibnamefont {Hilton}}, \bibinfo {author}
  {\bibfnamefont {K.}~\bibnamefont {Karimi}}, \bibinfo {author} {\bibfnamefont
  {E.}~\bibnamefont {Ladizinsky}}, \bibinfo {author} {\bibfnamefont
  {N.}~\bibnamefont {Ladizinsky}}, \bibinfo {author} {\bibfnamefont
  {T.}~\bibnamefont {Oh}}, \bibinfo {author} {\bibfnamefont {I.}~\bibnamefont
  {Perminov}}, \bibinfo {author} {\bibfnamefont {C.}~\bibnamefont {Rich}},
  \bibinfo {author} {\bibfnamefont {M.~C.}\ \bibnamefont {Thom}}, \bibinfo
  {author} {\bibfnamefont {E.}~\bibnamefont {Tolkacheva}}, \bibinfo {author}
  {\bibfnamefont {C.~J.~S.}\ \bibnamefont {Truncik}}, \bibinfo {author}
  {\bibfnamefont {S.}~\bibnamefont {Uchaikin}}, \bibinfo {author}
  {\bibfnamefont {J.}~\bibnamefont {Wang}}, \bibinfo {author} {\bibfnamefont
  {B.}~\bibnamefont {Wilson}}, \ and\ \bibinfo {author} {\bibfnamefont
  {G.}~\bibnamefont {Rose}},\ }\bibfield  {title} {\enquote {\bibinfo {title}
  {Quantum annealing with manufactured spins},}\ }\href {\doibase
  10.1038/nature10012} {\bibfield  {journal} {\bibinfo  {journal} {Nature}\
  }\textbf {\bibinfo {volume} {473}},\ \bibinfo {pages} {194--198} (\bibinfo
  {year} {2011})}\BibitemShut {NoStop}%
\bibitem [{\citenamefont {Hen}\ \emph {et~al.}(2015)\citenamefont {Hen},
  \citenamefont {Job}, \citenamefont {Albash}, \citenamefont {R{\o}nnow},
  \citenamefont {Troyer},\ and\ \citenamefont {Lidar}}]{Hen:2015rt}%
  \BibitemOpen
  \bibfield  {author} {\bibinfo {author} {\bibfnamefont {Itay}\ \bibnamefont
  {Hen}}, \bibinfo {author} {\bibfnamefont {Joshua}\ \bibnamefont {Job}},
  \bibinfo {author} {\bibfnamefont {Tameem}\ \bibnamefont {Albash}}, \bibinfo
  {author} {\bibfnamefont {Troels~F.}\ \bibnamefont {R{\o}nnow}}, \bibinfo
  {author} {\bibfnamefont {Matthias}\ \bibnamefont {Troyer}}, \ and\ \bibinfo
  {author} {\bibfnamefont {Daniel~A.}\ \bibnamefont {Lidar}},\ }\bibfield
  {title} {\enquote {\bibinfo {title} {Probing for quantum speedup in
  spin-glass problems with planted solutions},}\ }\href
  {http://link.aps.org/doi/10.1103/PhysRevA.92.042325} {\bibfield  {journal}
  {\bibinfo  {journal} {{Phys. Rev. A}}\ }\textbf {\bibinfo {volume} {92}},\
  \bibinfo {pages} {042325--} (\bibinfo {year} {2015})}\BibitemShut {NoStop}%
\bibitem [{\citenamefont {Boixo}\ \emph {et~al.}(2013)\citenamefont {Boixo},
  \citenamefont {Albash}, \citenamefont {Spedalieri}, \citenamefont
  {Chancellor},\ and\ \citenamefont {Lidar}}]{q-sig}%
  \BibitemOpen
  \bibfield  {author} {\bibinfo {author} {\bibfnamefont {Sergio}\ \bibnamefont
  {Boixo}}, \bibinfo {author} {\bibfnamefont {Tameem}\ \bibnamefont {Albash}},
  \bibinfo {author} {\bibfnamefont {Federico~M.}\ \bibnamefont {Spedalieri}},
  \bibinfo {author} {\bibfnamefont {Nicholas}\ \bibnamefont {Chancellor}}, \
  and\ \bibinfo {author} {\bibfnamefont {Daniel~A.}\ \bibnamefont {Lidar}},\
  }\bibfield  {title} {\enquote {\bibinfo {title} {Experimental signature of
  programmable quantum annealing},}\ }\href {\doibase 10.1038/ncomms3067}
  {\bibfield  {journal} {\bibinfo  {journal} {Nat. Commun.}\ }\textbf {\bibinfo
  {volume} {4}},\ \bibinfo {pages} {2067} (\bibinfo {year} {2013})}\BibitemShut
  {NoStop}%
\bibitem [{\citenamefont {Albash}\ \emph {et~al.}(2015)\citenamefont {Albash},
  \citenamefont {Vinci}, \citenamefont {Mishra}, \citenamefont {Warburton},\
  and\ \citenamefont {Lidar}}]{q-sig2}%
  \BibitemOpen
  \bibfield  {author} {\bibinfo {author} {\bibfnamefont {Tameem}\ \bibnamefont
  {Albash}}, \bibinfo {author} {\bibfnamefont {Walter}\ \bibnamefont {Vinci}},
  \bibinfo {author} {\bibfnamefont {Anurag}\ \bibnamefont {Mishra}}, \bibinfo
  {author} {\bibfnamefont {Paul~A.}\ \bibnamefont {Warburton}}, \ and\ \bibinfo
  {author} {\bibfnamefont {Daniel~A.}\ \bibnamefont {Lidar}},\ }\bibfield
  {title} {\enquote {\bibinfo {title} {Consistency tests of classical and
  quantum models for a quantum annealer},}\ }\href
  {http://link.aps.org/doi/10.1103/PhysRevA.91.042314} {\bibfield  {journal}
  {\bibinfo  {journal} {Phys. Rev. A}\ }\textbf {\bibinfo {volume} {91}},\
  \bibinfo {pages} {042314--} (\bibinfo {year} {2015})}\BibitemShut {NoStop}%
\bibitem [{QEO()}]{QEO}%
  \BibitemOpen
  \href {https://www.iarpa.gov/index.php/research-programs/qeo} {\enquote
  {\bibinfo {title} {Quantum enhanced optimization (qeo)},}\ }\BibitemShut
  {NoStop}%
\bibitem [{\citenamefont {Tolpygo}\ \emph
  {et~al.}(2015{\natexlab{a}})\citenamefont {Tolpygo}, \citenamefont
  {Bolkhovsky}, \citenamefont {Weir}, \citenamefont {Johnson}, \citenamefont
  {Gouker},\ and\ \citenamefont {Oliver}}]{LL1}%
  \BibitemOpen
  \bibfield  {author} {\bibinfo {author} {\bibfnamefont {S.~K.}\ \bibnamefont
  {Tolpygo}}, \bibinfo {author} {\bibfnamefont {V.}~\bibnamefont {Bolkhovsky}},
  \bibinfo {author} {\bibfnamefont {T.~J.}\ \bibnamefont {Weir}}, \bibinfo
  {author} {\bibfnamefont {L.~M.}\ \bibnamefont {Johnson}}, \bibinfo {author}
  {\bibfnamefont {M.~A.}\ \bibnamefont {Gouker}}, \ and\ \bibinfo {author}
  {\bibfnamefont {W.~D.}\ \bibnamefont {Oliver}},\ }\bibfield  {title}
  {\enquote {\bibinfo {title} {Fabrication process and properties of
  fully-planarized deep-submicron nb/al-alox/nb josephson junctions for vlsi
  circuits},}\ }\href {\doibase 10.1109/TASC.2014.2374836} {\bibfield
  {journal} {\bibinfo  {journal} {IEEE Transactions on Applied
  Superconductivity}\ }\textbf {\bibinfo {volume} {25}},\ \bibinfo {pages}
  {1--12} (\bibinfo {year} {2015}{\natexlab{a}})}\BibitemShut {NoStop}%
\bibitem [{\citenamefont {Tolpygo}\ \emph
  {et~al.}(2015{\natexlab{b}})\citenamefont {Tolpygo}, \citenamefont
  {Bolkhovsky}, \citenamefont {Weir}, \citenamefont {Galbraith}, \citenamefont
  {Johnson}, \citenamefont {Gouker},\ and\ \citenamefont {Semenov}}]{LL2}%
  \BibitemOpen
  \bibfield  {author} {\bibinfo {author} {\bibfnamefont {S.~K.}\ \bibnamefont
  {Tolpygo}}, \bibinfo {author} {\bibfnamefont {V.}~\bibnamefont {Bolkhovsky}},
  \bibinfo {author} {\bibfnamefont {T.~J.}\ \bibnamefont {Weir}}, \bibinfo
  {author} {\bibfnamefont {C.~J.}\ \bibnamefont {Galbraith}}, \bibinfo {author}
  {\bibfnamefont {L.~M.}\ \bibnamefont {Johnson}}, \bibinfo {author}
  {\bibfnamefont {M.~A.}\ \bibnamefont {Gouker}}, \ and\ \bibinfo {author}
  {\bibfnamefont {V.~K.}\ \bibnamefont {Semenov}},\ }\bibfield  {title}
  {\enquote {\bibinfo {title} {Inductance of circuit structures for mit ll
  superconductor electronics fabrication process with 8 niobium layers},}\
  }\href {\doibase 10.1109/TASC.2014.2369213} {\bibfield  {journal} {\bibinfo
  {journal} {IEEE Transactions on Applied Superconductivity}\ }\textbf
  {\bibinfo {volume} {25}},\ \bibinfo {pages} {1--5} (\bibinfo {year}
  {2015}{\natexlab{b}})}\BibitemShut {NoStop}%
\bibitem [{\citenamefont {Jin}\ \emph {et~al.}(2015)\citenamefont {Jin},
  \citenamefont {Kamal}, \citenamefont {Sears}, \citenamefont {Gudmundsen},
  \citenamefont {Hover}, \citenamefont {Miloshi}, \citenamefont {Slattery},
  \citenamefont {Yan}, \citenamefont {Yoder}, \citenamefont {Orlando},
  \citenamefont {Gustavsson},\ and\ \citenamefont {Oliver}}]{LL3}%
  \BibitemOpen
  \bibfield  {author} {\bibinfo {author} {\bibfnamefont {X.~Y.}\ \bibnamefont
  {Jin}}, \bibinfo {author} {\bibfnamefont {A.}~\bibnamefont {Kamal}}, \bibinfo
  {author} {\bibfnamefont {A.~P.}\ \bibnamefont {Sears}}, \bibinfo {author}
  {\bibfnamefont {T.}~\bibnamefont {Gudmundsen}}, \bibinfo {author}
  {\bibfnamefont {D.}~\bibnamefont {Hover}}, \bibinfo {author} {\bibfnamefont
  {J.}~\bibnamefont {Miloshi}}, \bibinfo {author} {\bibfnamefont
  {R.}~\bibnamefont {Slattery}}, \bibinfo {author} {\bibfnamefont
  {F.}~\bibnamefont {Yan}}, \bibinfo {author} {\bibfnamefont {J.}~\bibnamefont
  {Yoder}}, \bibinfo {author} {\bibfnamefont {T.~P.}\ \bibnamefont {Orlando}},
  \bibinfo {author} {\bibfnamefont {S.}~\bibnamefont {Gustavsson}}, \ and\
  \bibinfo {author} {\bibfnamefont {W.~D.}\ \bibnamefont {Oliver}},\ }\bibfield
   {title} {\enquote {\bibinfo {title} {Thermal and residual excited-state
  population in a 3d transmon qubit},}\ }\href {\doibase
  10.1103/PhysRevLett.114.240501} {\bibfield  {journal} {\bibinfo  {journal}
  {Phys. Rev. Lett.}\ }\textbf {\bibinfo {volume} {114}},\ \bibinfo {pages}
  {240501} (\bibinfo {year} {2015})}\BibitemShut {NoStop}%
\bibitem [{DW2()}]{DW2000}%
  \BibitemOpen
  \href
  {https://www.dwavesys.com/press-releases/d-wave-systems-previews-2000-qubit-quantum-system}
  {\enquote {\bibinfo {title} {D-wave systems previews 2000-qubit quantum
  system},}\ }\BibitemShut {NoStop}%
\bibitem [{\citenamefont {Young}\ \emph {et~al.}(2008)\citenamefont {Young},
  \citenamefont {Knysh},\ and\ \citenamefont {Smelyanskiy}}]{young:08}%
  \BibitemOpen
  \bibfield  {author} {\bibinfo {author} {\bibfnamefont {A.~P.}\ \bibnamefont
  {Young}}, \bibinfo {author} {\bibfnamefont {S.}~\bibnamefont {Knysh}}, \ and\
  \bibinfo {author} {\bibfnamefont {V.~N.}\ \bibnamefont {Smelyanskiy}},\
  }\bibfield  {title} {\enquote {\bibinfo {title} {Size dependence of the
  minimum excitation gap in the {Q}uantum {A}diabatic {A}lgorithm},}\
  }\href@noop {} {\bibfield  {journal} {\bibinfo  {journal} {Phys. Rev. Lett.}\
  }\textbf {\bibinfo {volume} {101}},\ \bibinfo {pages} {170503} (\bibinfo
  {year} {2008})},\ \Eprint {http://arxiv.org/abs/(arXiv:0803.3971)}
  {(arXiv:0803.3971)} \BibitemShut {NoStop}%
\bibitem [{\citenamefont {Young}\ \emph {et~al.}(2010)\citenamefont {Young},
  \citenamefont {Knysh},\ and\ \citenamefont {Smelyanskiy}}]{young:10}%
  \BibitemOpen
  \bibfield  {author} {\bibinfo {author} {\bibfnamefont {A.~P.}\ \bibnamefont
  {Young}}, \bibinfo {author} {\bibfnamefont {S.}~\bibnamefont {Knysh}}, \ and\
  \bibinfo {author} {\bibfnamefont {V.~N.}\ \bibnamefont {Smelyanskiy}},\
  }\bibfield  {title} {\enquote {\bibinfo {title} {First order phase transition
  in the {Q}uantum {A}diabatic {A}lgorithm},}\ }\href@noop {} {\bibfield
  {journal} {\bibinfo  {journal} {Phys. Rev. Lett.}\ }\textbf {\bibinfo
  {volume} {104}},\ \bibinfo {pages} {020502} (\bibinfo {year} {2010})},\
  \Eprint {http://arxiv.org/abs/(arXiv:0910.1378)} {(arXiv:0910.1378)}
  \BibitemShut {NoStop}%
\bibitem [{\citenamefont {Hen}\ and\ \citenamefont {Young}(2011)}]{hen:11}%
  \BibitemOpen
  \bibfield  {author} {\bibinfo {author} {\bibfnamefont {I.}~\bibnamefont
  {Hen}}\ and\ \bibinfo {author} {\bibfnamefont {A.~P.}\ \bibnamefont
  {Young}},\ }\bibfield  {title} {\enquote {\bibinfo {title} {Exponential
  complexity of the quantum adiabatic algorithm for certain satisfiability
  problems},}\ }\href@noop {} {\bibfield  {journal} {\bibinfo  {journal} {Phys.
  Rev. E.}\ }\textbf {\bibinfo {volume} {84}},\ \bibinfo {pages} {061152}
  (\bibinfo {year} {2011})},\ \Eprint {http://arxiv.org/abs/arXiv:1109.6872v2}
  {arXiv:1109.6872v2} \BibitemShut {NoStop}%
\bibitem [{\citenamefont {Hen}(2012)}]{hen:12}%
  \BibitemOpen
  \bibfield  {author} {\bibinfo {author} {\bibfnamefont {I.}~\bibnamefont
  {Hen}},\ }\bibfield  {title} {\enquote {\bibinfo {title} {Excitation gap from
  optimized correlation functions in quantum {M}onte {C}arlo simulations},}\
  }\href@noop {} {\bibfield  {journal} {\bibinfo  {journal} {Phys. Rev. E.}\
  }\textbf {\bibinfo {volume} {85}},\ \bibinfo {pages} {036705} (\bibinfo
  {year} {2012})},\ \Eprint {http://arxiv.org/abs/arXiv:1112.2269v2}
  {arXiv:1112.2269v2} \BibitemShut {NoStop}%
\bibitem [{\citenamefont {Farhi}\ \emph {et~al.}(2012)\citenamefont {Farhi},
  \citenamefont {Gosset}, \citenamefont {Hen}, \citenamefont {Sandvik},
  \citenamefont {Shor}, \citenamefont {Young},\ and\ \citenamefont
  {Zamponi}}]{farhi:12}%
  \BibitemOpen
  \bibfield  {author} {\bibinfo {author} {\bibfnamefont {E.}~\bibnamefont
  {Farhi}}, \bibinfo {author} {\bibfnamefont {D.}~\bibnamefont {Gosset}},
  \bibinfo {author} {\bibfnamefont {I.}~\bibnamefont {Hen}}, \bibinfo {author}
  {\bibfnamefont {A.~W.}\ \bibnamefont {Sandvik}}, \bibinfo {author}
  {\bibfnamefont {P.}~\bibnamefont {Shor}}, \bibinfo {author} {\bibfnamefont
  {A.~P.}\ \bibnamefont {Young}}, \ and\ \bibinfo {author} {\bibfnamefont
  {F.}~\bibnamefont {Zamponi}},\ }\bibfield  {title} {\enquote {\bibinfo
  {title} {Performance of the quantum adiabatic algorithm on random instances
  of two optimization problems on regular hypergraphs},}\ }\href@noop {}
  {\bibfield  {journal} {\bibinfo  {journal} {Phys. Rev. A}\ }\textbf {\bibinfo
  {volume} {86}},\ \bibinfo {pages} {052334} (\bibinfo {year}
  {2012})}\BibitemShut {NoStop}%
\bibitem [{\citenamefont {Roland}\ and\ \citenamefont
  {Cerf}(2002)}]{roland:02}%
  \BibitemOpen
  \bibfield  {author} {\bibinfo {author} {\bibfnamefont {J.}~\bibnamefont
  {Roland}}\ and\ \bibinfo {author} {\bibfnamefont {N.~J.}\ \bibnamefont
  {Cerf}},\ }\bibfield  {title} {\enquote {\bibinfo {title} {Quantum search by
  local adiabatic evolution},}\ }\href@noop {} {\bibfield  {journal} {\bibinfo
  {journal} {Phys. Rev. A}\ }\textbf {\bibinfo {volume} {65}},\ \bibinfo
  {pages} {042308} (\bibinfo {year} {2002})}\BibitemShut {NoStop}%
\bibitem [{\citenamefont {{Farhi}}\ and\ \citenamefont
  {{Gutmann}}(1996)}]{analogAnalogue}%
  \BibitemOpen
  \bibfield  {author} {\bibinfo {author} {\bibfnamefont {E.}~\bibnamefont
  {{Farhi}}}\ and\ \bibinfo {author} {\bibfnamefont {S.}~\bibnamefont
  {{Gutmann}}},\ }\bibfield  {title} {\enquote {\bibinfo {title} {{An Analog
  Analogue of a Digital Quantum Computation}},}\ }\href@noop {} {\bibfield
  {journal} {\bibinfo  {journal} {eprint arXiv:quant-ph/9612026}\ } (\bibinfo
  {year} {1996})},\ \Eprint {http://arxiv.org/abs/quant-ph/9612026}
  {quant-ph/9612026} \BibitemShut {NoStop}%
\bibitem [{\citenamefont {van Dam}\ \emph {et~al.}(2001)\citenamefont {van
  Dam}, \citenamefont {Mosca},\ and\ \citenamefont {Vazirani}}]{howPowerful}%
  \BibitemOpen
  \bibfield  {author} {\bibinfo {author} {\bibfnamefont {W.}~\bibnamefont {van
  Dam}}, \bibinfo {author} {\bibfnamefont {M.}~\bibnamefont {Mosca}}, \ and\
  \bibinfo {author} {\bibfnamefont {U.}~\bibnamefont {Vazirani}},\ }\bibfield
  {title} {\enquote {\bibinfo {title} {How powerful is adiabatic quantum
  computation?}}\ }in\ \href {\doibase 10.1109/SFCS.2001.959902} {\emph
  {\bibinfo {booktitle} {Proceedings 2001 IEEE International Conference on
  Cluster Computing}}}\ (\bibinfo {year} {2001})\ pp.\ \bibinfo {pages}
  {279--287}\BibitemShut {NoStop}%
\bibitem [{\citenamefont {Jansen}\ \emph {et~al.}(2007)\citenamefont {Jansen},
  \citenamefont {Ruskai},\ and\ \citenamefont {Seiler}}]{jansen:07}%
  \BibitemOpen
  \bibfield  {author} {\bibinfo {author} {\bibfnamefont {S.}~\bibnamefont
  {Jansen}}, \bibinfo {author} {\bibfnamefont {M.~B.}\ \bibnamefont {Ruskai}},
  \ and\ \bibinfo {author} {\bibfnamefont {R.}~\bibnamefont {Seiler}},\
  }\bibfield  {title} {\enquote {\bibinfo {title} {Bounds for the adiabatic
  approximation with applications to quantum computation},}\ }\href@noop {}
  {\bibfield  {journal} {\bibinfo  {journal} {J. Math. Phys.}\ }\textbf
  {\bibinfo {volume} {47}},\ \bibinfo {pages} {102111} (\bibinfo {year}
  {2007})}\BibitemShut {NoStop}%
\bibitem [{\citenamefont {Lidar}\ \emph {et~al.}(2009)\citenamefont {Lidar},
  \citenamefont {Rezakhani},\ and\ \citenamefont {Hamma}}]{lidarGap}%
  \BibitemOpen
  \bibfield  {author} {\bibinfo {author} {\bibfnamefont {D.~A.}\ \bibnamefont
  {Lidar}}, \bibinfo {author} {\bibfnamefont {A.~T.}\ \bibnamefont
  {Rezakhani}}, \ and\ \bibinfo {author} {\bibfnamefont {A.}~\bibnamefont
  {Hamma}},\ }\bibfield  {title} {\enquote {\bibinfo {title} {Adiabatic
  approximation with exponential accuracy for many-body systems and quantum
  computation},}\ }\href {\doibase http://dx.doi.org/10.1063/1.3236685}
  {\bibfield  {journal} {\bibinfo  {journal} {J. Math. Phys.}\ }\textbf
  {\bibinfo {volume} {50}},\ \bibinfo {eid} {102106} (\bibinfo {year}
  {2009})}\BibitemShut {NoStop}%
\bibitem [{\citenamefont {Kato}(1951)}]{kato:51}%
  \BibitemOpen
  \bibfield  {author} {\bibinfo {author} {\bibfnamefont {T.}~\bibnamefont
  {Kato}},\ }\bibfield  {title} {\enquote {\bibinfo {title} {On the adiabatic
  theorem of quantum mechanics},}\ }\href@noop {} {\bibfield  {journal}
  {\bibinfo  {journal} {J. Phys. Soc. Jap.}\ }\textbf {\bibinfo {volume} {5}},\
  \bibinfo {pages} {435} (\bibinfo {year} {1951})}\BibitemShut {NoStop}%
\bibitem [{\citenamefont {Hen}(2014{\natexlab{a}})}]{hen:14}%
  \BibitemOpen
  \bibfield  {author} {\bibinfo {author} {\bibfnamefont {I.}~\bibnamefont
  {Hen}},\ }\bibfield  {title} {\enquote {\bibinfo {title} {Continuous-time
  quantum algorithms for unstructured problems},}\ }\href@noop {} {\bibfield
  {journal} {\bibinfo  {journal} {J. Phys. A: Math. Theor.}\ }\textbf {\bibinfo
  {volume} {47}},\ \bibinfo {pages} {045305} (\bibinfo {year}
  {2014}{\natexlab{a}})},\ \Eprint {http://arxiv.org/abs/arXiv:1302.7256}
  {arXiv:1302.7256} \BibitemShut {NoStop}%
\bibitem [{\citenamefont {Hen}(2014{\natexlab{b}})}]{hen:14b}%
  \BibitemOpen
  \bibfield  {author} {\bibinfo {author} {\bibfnamefont {Itay}\ \bibnamefont
  {Hen}},\ }\bibfield  {title} {\enquote {\bibinfo {title} {How fast can
  quantum annealers count?}}\ }\href
  {http://stacks.iop.org/1751-8121/47/i=23/a=235304} {\bibfield  {journal}
  {\bibinfo  {journal} {Journal of Physics A: Mathematical and Theoretical}\
  }\textbf {\bibinfo {volume} {47}},\ \bibinfo {pages} {235304} (\bibinfo
  {year} {2014}{\natexlab{b}})}\BibitemShut {NoStop}%
\bibitem [{\citenamefont {Hen}(2017)}]{realizableAQCsearch}%
  \BibitemOpen
  \bibfield  {author} {\bibinfo {author} {\bibfnamefont {I.}~\bibnamefont
  {Hen}},\ }\bibfield  {title} {\enquote {\bibinfo {title} {Realizable quantum
  adiabatic search},}\ }\href {http://stacks.iop.org/0295-5075/118/i=3/a=30003}
  {\bibfield  {journal} {\bibinfo  {journal} {EPL (Europhysics Letters)}\
  }\textbf {\bibinfo {volume} {118}},\ \bibinfo {pages} {30003} (\bibinfo
  {year} {2017})}\BibitemShut {NoStop}%
\bibitem [{\citenamefont {Roland}\ and\ \citenamefont
  {Cerf}(2003)}]{roland:03}%
  \BibitemOpen
  \bibfield  {author} {\bibinfo {author} {\bibfnamefont {J.}~\bibnamefont
  {Roland}}\ and\ \bibinfo {author} {\bibfnamefont {N.~J.}\ \bibnamefont
  {Cerf}},\ }\bibfield  {title} {\enquote {\bibinfo {title} {Adiabatic quantum
  search algorithm for structured problems},}\ }\href@noop {} {\bibfield
  {journal} {\bibinfo  {journal} {Phys. Rev. A}\ }\textbf {\bibinfo {volume}
  {68}},\ \bibinfo {pages} {062312} (\bibinfo {year} {2003})}\BibitemShut
  {NoStop}%
\bibitem [{\citenamefont {Childs}\ and\ \citenamefont
  {Goldstone}(2004)}]{childsGoldstone}%
  \BibitemOpen
  \bibfield  {author} {\bibinfo {author} {\bibfnamefont {Andrew~M.}\
  \bibnamefont {Childs}}\ and\ \bibinfo {author} {\bibfnamefont {Jeffrey}\
  \bibnamefont {Goldstone}},\ }\bibfield  {title} {\enquote {\bibinfo {title}
  {Spatial search by quantum walk},}\ }\href {\doibase
  10.1103/PhysRevA.70.022314} {\bibfield  {journal} {\bibinfo  {journal} {Phys.
  Rev. A}\ }\textbf {\bibinfo {volume} {70}},\ \bibinfo {pages} {022314}
  (\bibinfo {year} {2004})}\BibitemShut {NoStop}%
\bibitem [{\citenamefont {Albash}\ and\ \citenamefont
  {Lidar}(2018)}]{RevModPhys.90.015002}%
  \BibitemOpen
  \bibfield  {author} {\bibinfo {author} {\bibfnamefont {Tameem}\ \bibnamefont
  {Albash}}\ and\ \bibinfo {author} {\bibfnamefont {Daniel~A.}\ \bibnamefont
  {Lidar}},\ }\bibfield  {title} {\enquote {\bibinfo {title} {Adiabatic quantum
  computation},}\ }\href {\doibase 10.1103/RevModPhys.90.015002} {\bibfield
  {journal} {\bibinfo  {journal} {Rev. Mod. Phys.}\ }\textbf {\bibinfo {volume}
  {90}},\ \bibinfo {pages} {015002} (\bibinfo {year} {2018})}\BibitemShut
  {NoStop}%
\bibitem [{\citenamefont {Jiang}\ \emph {et~al.}(2017)\citenamefont {Jiang},
  \citenamefont {Rieffel},\ and\ \citenamefont {Wang}}]{QAOA}%
  \BibitemOpen
  \bibfield  {author} {\bibinfo {author} {\bibfnamefont {Zhang}\ \bibnamefont
  {Jiang}}, \bibinfo {author} {\bibfnamefont {Eleanor~G.}\ \bibnamefont
  {Rieffel}}, \ and\ \bibinfo {author} {\bibfnamefont {Zhihui}\ \bibnamefont
  {Wang}},\ }\bibfield  {title} {\enquote {\bibinfo {title} {Near-optimal
  quantum circuit for grover's unstructured search using a transverse field},}\
  }\href {\doibase 10.1103/PhysRevA.95.062317} {\bibfield  {journal} {\bibinfo
  {journal} {Phys. Rev. A}\ }\textbf {\bibinfo {volume} {95}},\ \bibinfo
  {pages} {062317} (\bibinfo {year} {2017})}\BibitemShut {NoStop}%
\bibitem [{\citenamefont {Aaronson}(2005)}]{Aaronson}%
  \BibitemOpen
  \bibfield  {author} {\bibinfo {author} {\bibfnamefont {Scott}\ \bibnamefont
  {Aaronson}},\ }\bibfield  {title} {\enquote {\bibinfo {title} {Guest column:
  Np-complete problems and physical reality},}\ }\href {\doibase
  10.1145/1052796.1052804} {\bibfield  {journal} {\bibinfo  {journal} {SIGACT
  News}\ }\textbf {\bibinfo {volume} {36}},\ \bibinfo {pages} {30--52}
  (\bibinfo {year} {2005})}\BibitemShut {NoStop}%
\bibitem [{\citenamefont {Vergis}\ \emph {et~al.}(1986)\citenamefont {Vergis},
  \citenamefont {Steiglitz},\ and\ \citenamefont {Dickinson}}]{VERGIS198691}%
  \BibitemOpen
  \bibfield  {author} {\bibinfo {author} {\bibfnamefont {Anastasios}\
  \bibnamefont {Vergis}}, \bibinfo {author} {\bibfnamefont {Kenneth}\
  \bibnamefont {Steiglitz}}, \ and\ \bibinfo {author} {\bibfnamefont {Bradley}\
  \bibnamefont {Dickinson}},\ }\bibfield  {title} {\enquote {\bibinfo {title}
  {The complexity of analog computation},}\ }\href {\doibase
  https://doi.org/10.1016/0378-4754(86)90105-9} {\bibfield  {journal} {\bibinfo
   {journal} {Mathematics and Computers in Simulation}\ }\textbf {\bibinfo
  {volume} {28}},\ \bibinfo {pages} {91 -- 113} (\bibinfo {year}
  {1986})}\BibitemShut {NoStop}%
\bibitem [{\citenamefont {Jackson}(1960)}]{analogComputing}%
  \BibitemOpen
  \bibfield  {author} {\bibinfo {author} {\bibfnamefont {Albert~S.}\
  \bibnamefont {Jackson}},\ }\href@noop {} {\emph {\bibinfo {title} {Analog
  Computation}}}\ (\bibinfo  {publisher} {McGraw-Hill},\ \bibinfo {address}
  {New York, USA},\ \bibinfo {year} {1960})\BibitemShut {NoStop}%
\bibitem [{\citenamefont {{Albash}}\ \emph {et~al.}(2019)\citenamefont
  {{Albash}}, \citenamefont {{Martin-Mayor}},\ and\ \citenamefont
  {{Hen}}}]{analogErrors}%
  \BibitemOpen
  \bibfield  {author} {\bibinfo {author} {\bibfnamefont {T.}~\bibnamefont
  {{Albash}}}, \bibinfo {author} {\bibfnamefont {V.}~\bibnamefont
  {{Martin-Mayor}}}, \ and\ \bibinfo {author} {\bibfnamefont {I.}~\bibnamefont
  {{Hen}}},\ }\bibfield  {title} {\enquote {\bibinfo {title} {{Analog Errors in
  Ising Machines}},}\ }\href@noop {} {\bibfield  {journal} {\bibinfo  {journal}
  {accepted for publication in Quantum Science \& Technology}\ } (\bibinfo
  {year} {2019})},\ \Eprint {http://arxiv.org/abs/1806.03744} {arXiv:1806.03744
  [quant-ph]} \BibitemShut {NoStop}%
\bibitem [{Note1()}]{Note1}%
  \BibitemOpen
  \bibinfo {note} {The Lagrangian can be interpreted as a system of rotors with
  time-varying potential and moments of inertia.}\BibitemShut {Stop}%
\bibitem [{Note2()}]{Note2}%
  \BibitemOpen
  \bibinfo {note} {One could consider a small addition to the kinetic term that
  would remove the degeneracy of its minima and enforce the $\theta _i=\pi /2$
  condition.}\BibitemShut {Stop}%
\bibitem [{Note3()}]{Note3}%
  \BibitemOpen
  \bibinfo {note} {In this case, the algorithm is reminiscent of the analog
  quantum search algorithm proposed in Ref.~\cite
  {analogAnalogue}.}\BibitemShut {Stop}%
\bibitem [{\citenamefont {Jonckheere}\ \emph {et~al.}(2013)\citenamefont
  {Jonckheere}, \citenamefont {Rezakhani},\ and\ \citenamefont
  {Ahmad}}]{Jonckheere2013}%
  \BibitemOpen
  \bibfield  {author} {\bibinfo {author} {\bibfnamefont {Edmond~A.}\
  \bibnamefont {Jonckheere}}, \bibinfo {author} {\bibfnamefont {Ali~T.}\
  \bibnamefont {Rezakhani}}, \ and\ \bibinfo {author} {\bibfnamefont {Farooq}\
  \bibnamefont {Ahmad}},\ }\bibfield  {title} {\enquote {\bibinfo {title}
  {Differential topology of adiabatically controlled quantum processes},}\
  }\href {\doibase 10.1007/s11128-012-0445-0} {\bibfield  {journal} {\bibinfo
  {journal} {Quantum Information Processing}\ }\textbf {\bibinfo {volume}
  {12}},\ \bibinfo {pages} {1515--1538} (\bibinfo {year} {2013})}\BibitemShut
  {NoStop}%
\bibitem [{\citenamefont {{Slutskii}}\ \emph {et~al.}(2019)\citenamefont
  {{Slutskii}}, \citenamefont {{Albash}}, \citenamefont {{Barash}},\ and\
  \citenamefont {{Hen}}}]{analogQAUS}%
  \BibitemOpen
  \bibfield  {author} {\bibinfo {author} {\bibfnamefont {Mikhail}\ \bibnamefont
  {{Slutskii}}}, \bibinfo {author} {\bibfnamefont {Tameem}\ \bibnamefont
  {{Albash}}}, \bibinfo {author} {\bibfnamefont {Lev}\ \bibnamefont
  {{Barash}}}, \ and\ \bibinfo {author} {\bibfnamefont {Itay}\ \bibnamefont
  {{Hen}}},\ }\bibfield  {title} {\enquote {\bibinfo {title} {{Analog Nature of
  Quantum Adiabatic Unstructured Search}},}\ }\href@noop {} {\bibfield
  {journal} {\bibinfo  {journal} {arXiv e-prints}\ ,\ \bibinfo {eid}
  {arXiv:1904.04420}} (\bibinfo {year} {2019})},\ \Eprint
  {http://arxiv.org/abs/1904.04420} {arXiv:1904.04420 [quant-ph]} \BibitemShut
  {NoStop}%
\bibitem [{\citenamefont {Berry}\ \emph
  {et~al.}(2014{\natexlab{a}})\citenamefont {Berry}, \citenamefont {Cleve},\
  and\ \citenamefont {Gharibian}}]{Berry}%
  \BibitemOpen
  \bibfield  {author} {\bibinfo {author} {\bibfnamefont {Dominic~W.}\
  \bibnamefont {Berry}}, \bibinfo {author} {\bibfnamefont {Richard}\
  \bibnamefont {Cleve}}, \ and\ \bibinfo {author} {\bibfnamefont {Sevag}\
  \bibnamefont {Gharibian}},\ }\bibfield  {title} {\enquote {\bibinfo {title}
  {Gate-efficient discrete simulations of continuous-time quantum query
  algorithms},}\ }\href {http://dl.acm.org/citation.cfm?id=2600498.2600499}
  {\bibfield  {journal} {\bibinfo  {journal} {Quantum Info. Comput.}\ }\textbf
  {\bibinfo {volume} {14}},\ \bibinfo {pages} {1--30} (\bibinfo {year}
  {2014}{\natexlab{a}})}\BibitemShut {NoStop}%
\bibitem [{\citenamefont {Berry}\ \emph
  {et~al.}(2014{\natexlab{b}})\citenamefont {Berry}, \citenamefont {Childs},
  \citenamefont {Cleve}, \citenamefont {Kothari},\ and\ \citenamefont
  {Somma}}]{Somma1}%
  \BibitemOpen
  \bibfield  {author} {\bibinfo {author} {\bibfnamefont {Dominic~W.}\
  \bibnamefont {Berry}}, \bibinfo {author} {\bibfnamefont {Andrew~M.}\
  \bibnamefont {Childs}}, \bibinfo {author} {\bibfnamefont {Richard}\
  \bibnamefont {Cleve}}, \bibinfo {author} {\bibfnamefont {Robin}\ \bibnamefont
  {Kothari}}, \ and\ \bibinfo {author} {\bibfnamefont {Rolando~D.}\
  \bibnamefont {Somma}},\ }\bibfield  {title} {\enquote {\bibinfo {title}
  {Exponential improvement in precision for simulating sparse hamiltonians},}\
  }in\ \href {\doibase 10.1145/2591796.2591854} {\emph {\bibinfo {booktitle}
  {Proceedings of the Forty-sixth Annual ACM Symposium on Theory of
  Computing}}},\ \bibinfo {series and number} {STOC '14}\ (\bibinfo
  {publisher} {ACM},\ \bibinfo {address} {New York, NY, USA},\ \bibinfo {year}
  {2014})\ pp.\ \bibinfo {pages} {283--292}\BibitemShut {NoStop}%
\bibitem [{\citenamefont {Berry}\ \emph {et~al.}(2015)\citenamefont {Berry},
  \citenamefont {Childs}, \citenamefont {Cleve}, \citenamefont {Kothari},\ and\
  \citenamefont {Somma}}]{Somma2}%
  \BibitemOpen
  \bibfield  {author} {\bibinfo {author} {\bibfnamefont {Dominic~W.}\
  \bibnamefont {Berry}}, \bibinfo {author} {\bibfnamefont {Andrew~M.}\
  \bibnamefont {Childs}}, \bibinfo {author} {\bibfnamefont {Richard}\
  \bibnamefont {Cleve}}, \bibinfo {author} {\bibfnamefont {Robin}\ \bibnamefont
  {Kothari}}, \ and\ \bibinfo {author} {\bibfnamefont {Rolando~D.}\
  \bibnamefont {Somma}},\ }\bibfield  {title} {\enquote {\bibinfo {title}
  {Simulating hamiltonian dynamics with a truncated taylor series},}\ }\href
  {\doibase 10.1103/PhysRevLett.114.090502} {\bibfield  {journal} {\bibinfo
  {journal} {Phys. Rev. Lett.}\ }\textbf {\bibinfo {volume} {114}},\ \bibinfo
  {pages} {090502} (\bibinfo {year} {2015})}\BibitemShut {NoStop}%
\bibitem [{\citenamefont {Cleve}\ \emph {et~al.}(2009)\citenamefont {Cleve},
  \citenamefont {Gottesman}, \citenamefont {Mosca}, \citenamefont {Somma},\
  and\ \citenamefont {Yonge-Mallo}}]{Somma3}%
  \BibitemOpen
  \bibfield  {author} {\bibinfo {author} {\bibfnamefont {Richard}\ \bibnamefont
  {Cleve}}, \bibinfo {author} {\bibfnamefont {Daniel}\ \bibnamefont
  {Gottesman}}, \bibinfo {author} {\bibfnamefont {Michele}\ \bibnamefont
  {Mosca}}, \bibinfo {author} {\bibfnamefont {Rolando~D.}\ \bibnamefont
  {Somma}}, \ and\ \bibinfo {author} {\bibfnamefont {David}\ \bibnamefont
  {Yonge-Mallo}},\ }\bibfield  {title} {\enquote {\bibinfo {title} {Efficient
  discrete-time simulations of continuous-time quantum query algorithms},}\
  }in\ \href {\doibase 10.1145/1536414.1536471} {\emph {\bibinfo {booktitle}
  {Proceedings of the Forty-first Annual ACM Symposium on Theory of
  Computing}}},\ \bibinfo {series and number} {STOC '09}\ (\bibinfo
  {publisher} {ACM},\ \bibinfo {address} {New York, NY, USA},\ \bibinfo {year}
  {2009})\ pp.\ \bibinfo {pages} {409--416}\BibitemShut {NoStop}%
\bibitem [{\citenamefont {Nielsen}\ and\ \citenamefont
  {Chuang}(2000)}]{nielsen:00}%
  \BibitemOpen
  \bibfield  {author} {\bibinfo {author} {\bibfnamefont {M.~A.}\ \bibnamefont
  {Nielsen}}\ and\ \bibinfo {author} {\bibfnamefont {I.~L.}\ \bibnamefont
  {Chuang}},\ }\href@noop {} {\emph {\bibinfo {title} {Quantum Computation and
  Quantum Information}}}\ (\bibinfo  {publisher} {Cambridge University Press},\
  \bibinfo {address} {Cambridge, England},\ \bibinfo {year} {2000})\BibitemShut
  {NoStop}%
\end{thebibliography}%
\end{document}